\documentclass[aps,prb,twocolumn,groupdaddress]{revtex4-1}

\usepackage{amsmath}
\usepackage{graphicx}

\usepackage{hyperref}
\hypersetup{colorlinks=true, citecolor=blue, urlcolor=blue, linkcolor=blue}

\newcommand{\bk}{{\bf k}}
\newcommand{\bK}{{\bf K}}
\newcommand{\bR}{{\bf R}}
\newcommand{\wn}{{i\omega_n}}

\begin{document}


\title{Continuous momentum dependence in the dynamical cluster approximation}


\author{Urs R. H\"ahner}
\email[]{haehneru@itp.phys.ethz.ch}
\affiliation{Institute for Theoretical Physics, ETH Zurich, 8093 Zurich, Switzerland}

\author{Thomas A. Maier}
\affiliation{Computational Science and Engineering Division, Center for Nanophase Materials Sciences, Oak Ridge National Laboratory, Oak Ridge, Tennessee 37831, USA}

\author{Thomas C. Schulthess}
\affiliation{Institute for Theoretical Physics, ETH Zurich, 8093 Zurich, Switzerland}
\affiliation{Swiss National Supercomputing Center, ETH Zurich, 6900 Lugano, Switzerland}


\date{\today}

\begin{abstract}
The dynamical cluster approximation~(DCA) is a quantum cluster extension to the single-site dynamical mean-field theory that incorporates spatially nonlocal dynamic correlations systematically and nonperturbatively.
The DCA$^+$ algorithm addresses the cluster shape dependence of the DCA and improves the convergence with cluster size by introducing a lattice self-energy with continuous momentum dependence.
However, we show that the DCA$^+$ algorithm is plagued by a fundamental problem when its self-consistency equations are formulated using the bare Green's function of the cluster.
This problem is most severe in the strongly correlated regime at low doping, where the DCA$^+$ self-energy becomes overly metallic and local, and persists to cluster sizes where the standard DCA has long converged.
In view of the failure of the DCA$^+$ algorithm, we propose to complement DCA simulations with a post-interpolation procedure for single-particle and two-particle correlation functions to preserve continuous momentum dependence and the associated benefits in the DCA.
We demonstrate the effectiveness of this practical approach with results for the half-filled and hole-doped two-dimensional Hubbard model.
\end{abstract}


\maketitle

\section{Introduction}
\label{sec:introduction}

The dynamical mean-field theory~(DMFT)~\cite{Georges:1996hv} provides a computationally feasible approach to simulating lattice models of correlated electron systems such as the Hubbard model.
Based on the assumption that correlations are local in space, the DMFT reduces the complexity of the infinite lattice model by mapping it to a single-site impurity problem.
The DMFT reaches its limitations whenever spatially nonlocal correlations between electrons play a dominant role in the properties of a material.
For example, this is the case in the copper-oxide based materials (cuprates), where a nonlocal pairing mechanism is believed to give rise to high-temperature superconductivity~\cite{Scalapino:2012ek}.
To understand such material behavior and eliminate the shortcomings of the single-site theory, several methods have been developed that extend the DMFT and systematically include spatial correlations.
These methods can be classified into two groups:
diagrammatic extensions~\cite{Rohringer:2018fk}, which add spatial correlations on all length scales with a perturbative expansion around the DMFT solution, and
quantum cluster extensions~\cite{Maier:2005do}, which capture short-range spatial correlations exactly by replacing the single-site impurity with a finite-size cluster.
In this paper, we discuss the second group.

The requirements for a satisfactory quantum cluster extension were specified by Gonis~\cite{Gonis:1992ts} in the context of disordered systems and reformulated by Jarrell~et~al.~\cite{Jarrell:2001jj} for ordered correlated systems:
\begin{enumerate}
    \item A quantum cluster theory for ordered correlated systems should include nonlocal fluctuations in a self-consistent way.
    \item It should yield the DMFT when the cluster size $N_c$ is one.
    \item It should become exact in the thermodynamic limit (i.e., $N_c \rightarrow \infty$).
    \item It should be causal in the sense that the Green's function and the self-energy of the cluster are analytic functions in the upper half-plane.
    \item It should obey the point group symmetry of the lattice.
    \item It should preserve translational invariance.
    \item It should be numerically practicable with respect to the complexity of the implementation and the computational costs.
\end{enumerate}

The dynamical cluster approximation~(DCA)~\cite{Hettler:1998jj, Hettler:2000es, Maier:2005do} is a quantum cluster extension of the DMFT that satisfies all of these requirements~\cite{Jarrell:2001jj}.
In the DCA the infinite lattice model is replaced by a finite-size cluster with periodic boundary conditions that is embedded in a self-consistent mean-field.
The reduction to an effective cluster problem is achieved by coarse-graining the Green's function in momentum space with a piecewise constant approximation of the lattice self-energy.

Another widely-used quantum cluster extension is the cellular DMFT~(CDMFT)~\cite{Kotliar:2001iy}.
In contrast to the DCA, the CDMFT is formulated in real space and uses a cluster with open boundary conditions.
Therefore, it lacks the requirement of translational invariance.
Without translational invariance, the CDMFT is also computationally more expensive, since quantities such as Green's functions depend one two, instead of one, cluster momenta or sites.
For a more detailed comparison of the CDMFT and the DCA, we refer the reader to the review article by Maier~et~al.~\cite{Maier:2005do}.

While the DCA meets all listed requirements, the piecewise constant approximation of the DCA lattice self-energy comes with the inherent drawbacks of jump discontinuities and coarse momentum resolution, strong cluster shape dependence, and uncontrolled convergence with cluster size.
Moreover, the fermionic sign problem of the underlying quantum Monte Carlo~(QMC) algorithm that is used to solve the effective cluster problem and limited computational resources often prevent access to large clusters.
Therefore, to examine finite-size effects, clusters are used that do not fulfill the point group symmetry of the lattice, in violation to one of the requirements.

To address the cluster shape dependence, improve the convergence with cluster size, and restore the symmetry of the lattice, Staar~et~al. developed the dynamical cluster approximation with continuous lattice self-energy~(DCA$^+$)~\cite{Staar:2013ec}.
In contrast to the DCA, the DCA$^+$ algorithm employs a lattice self-energy with continuous momentum dependence in the coarse-graining of the Green's function.
This continuous lattice self-energy is determined through a deconvolution of the interpolated cluster self-energy.
As a side effect, the DCA$^+$ algorithm, apparently, is also less affected by the fermionic sign problem.

Staar~et~al. mentioned in the original paper~\cite{Staar:2013ec} that self-consistency in the DCA$^+$ algorithm requires the self-energy to be localized on the cluster so that the lattice mapping, i.e., the interpolation of the cluster self-energy and subsequent deconvolution to determine the lattice self-energy, converges.
Consequently, failure of the DCA$^+$ algorithm is to be expected for small clusters in the strongly correlated regime at low doping, where correlations are long-ranged.
In accordance with this presumption, Vu{\v c}i{\v c}evi{\'c}~et~al.~\cite{Vucicevic:2018ev} reported failure in the aforementioned part of the phase diagram in terms of a far too metallic and local self-energy and linked it to causality violations they observed in the DCA$^+$ hybridization function, which describes the coupling of the cluster to the mean-field host.

The focus of this paper is twofold.
First, we want to uncover the fundamental problem of the DCA$^+$ algorithm that leads to the observed failure in the strongly correlated regime at low doping.
We will show that the failure exists even in the large cluster limit.
Second, we will argue that the interpolation procedure introduced by Staar~et~al.~\cite{Staar:2013ec} for the DCA$^+$ algorithm can be applied ex post to DCA calculations to preserve many benefits of DCA$^+$ over the standard DCA without being affected by the described problem.

All analysis is done for the two-dimensional single-band Hubbard model on the square lattice, described by the Hamiltonian
\begin{equation}
    H = -t \sum_{\langle i, j \rangle, \sigma} c_{i \sigma}^\dagger c_{j \sigma} + U \sum_i n_{i \uparrow} n_{i \downarrow} ,
\end{equation}
where $t$ is the nearest-neighbor hopping amplitude and $U$ the on-site Coulomb interaction.
The operator $c_{i \sigma}^\dagger (c_{i \sigma})$ creates (annihilates) an electron with spin $\sigma$ at site $i$, and $n_{i \sigma} = c_{i \sigma}^\dagger c_{i \sigma}$ is the corresponding number operator.

The remainder of this paper is organized as follows:
Section~\ref{sec:methods} reviews the DCA and DCA$^+$ algorithms,
Sec.~\ref{sec:failure_DCA+} discusses the fundamental problem in the DCA$^+$ algorithm,
Sec.~\ref{sec:DCA_with_post_interpolation_application} presents a practical approach to preserve continuous momentum dependence in the DCA, and
Sec.~\ref{sec:conclusions} summarizes the results and discusses their consequences.

\section{Methods}
\label{sec:methods}

\subsection{Dynamical cluster approximation}

The DCA algorithm is formulated in reciprocal space.
To reduce the infinite lattice to a finite size cluster of $N_c$ sites, the first Brillouin zone~(BZ) is discretized into $N_c$ cluster momenta $\left\{\bK\right\}$.
To coarse-grain the degrees of freedom not represented by the cluster, the volume of the first Brillouin zone is divided into a set of patches, each of which is centered around a unique cluster momentum $\bK$.
Integration over the patches is performed by introducing the patch functions
\begin{equation}
\label{eq:patch_function}
    \phi_\bK(\bk) =
        \left\{\begin{array}{ll}
            1, & \bk \in \bK^\text{th} \text{ patch} , \\
            0, & \text{otherwise} .
        \end{array}\right.
\end{equation}

The main assumption in the DCA is that correlations are sufficiently short-ranged.
Accordingly, the lattice self-energy $\Sigma(\bk, \wn)$ is assumed to be only weakly momentum dependent and well approximated by a piecewise constant continuation of the cluster self-energy $\Sigma_c(\bK, \wn)$ in terms of the patch functions $\left\{\phi_\bK(\bk)\right\}$,
\begin{equation}
\label{eq:Sigma_DCA}
    \Sigma(\bk, \wn) \approx \Sigma_\text{DCA}(\bk, \wn) = \sum_\bK \phi_\bK(\bk) \Sigma_c(\bK, \wn) .
\end{equation}
In contrast to finite-size methods, where the cluster is solved in isolation, the DCA defines an effective cluster problem by coarse-graining the lattice Green's function using approximation~(\ref{eq:Sigma_DCA}),
\begin{align}
    \bar{G}(\bK, \wn) = \frac{N_c}{V_\text{BZ}}
        &\int_\text{BZ} \! d\bk \, \phi_\bK(\bk) \nonumber \\
        &\times \left[ \wn + \mu - \epsilon_\bk - \Sigma_\text{DCA}(\bk, \wn) \right]^{-1} . \label{eq:G_coarsegrained}
\end{align}
The self-energy $\Sigma_c(\bK, \wn)$ of the effective cluster problem is obtained by taking the functional derivative of the Luttinger-Ward functional $\Phi$~\cite{Luttinger:1960ht} with respect to the coarse-grained Green's function $\bar{G}(\bK, \wn)$~\cite{Baym:1961do},
\begin{equation}
\label{eq:new_Sigma_cluster}
    \Sigma_c(\bK, \wn) = \frac{\delta \Phi[\bar{G}(\bK, \wn)]}{\delta \bar{G}(\bK, \wn)} .
\end{equation}
Eqs.~(\ref{eq:Sigma_DCA}), (\ref{eq:G_coarsegrained}), and (\ref{eq:new_Sigma_cluster}) form the DCA self-consistency equations, which are iterated until convergence.
We will refer to this form of the self-consistency loop as the $\Phi$-\emph{formulation}.

State-of-the-art cluster solvers such as continuous-time QMC methods~\cite{Gull:2011jd} compute the new cluster self-energy $\Sigma_c(\bK, \wn)$ in a bare expansion in terms of the bare Green's function instead of a skeleton expansion with the interacting Green's function, as required by Eq.~(\ref{eq:new_Sigma_cluster}).
Hence, to prevent overcounting of self-energy diagrams, these methods need to be set up with the bare Green's function $\mathcal{G}_0(\bK, \wn)$ of the effective cluster problem defined by the coarse-grained Green's function $\bar{G}(\bK, \wn)$.
$\mathcal{G}_0(\bK, \wn)$ is computed from a reversed Dyson equation,
\begin{equation}
\label{eq:cluster-exclusion}
    \mathcal{G}_0(\bK, \wn) = \left[ \bar{G}^{-1}(\bK, \wn) + \Sigma_c(\bK, \wn) \right]^{-1} .
\end{equation}
The last equation, also called \emph{cluster-exclusion step} as it removes the self-energy contributions on the cluster, will be shown to play a crucial role in the failure of the DCA$^+$ algorithm.
We will refer to this modified self-consistency loop as the $\mathcal{G}_0$-\emph{formulation}.

\subsection{The DCA$^+$ algorithm}

As outlined in Sec.~\ref{sec:introduction}, the DCA suffers from problems due to the piecewise constant approximation of the lattice self-energy, Eq.~(\ref{eq:Sigma_DCA}).
The goal of the DCA$^+$ algorithm is to cure these problems by introducing a smooth lattice self-energy.
The primary insight is that Eq.~(\ref{eq:Sigma_DCA}) can be inverted and, as a result, yields a coarse-graining equation for the lattice self-energy,
\begin{equation}
\label{eq:self-energy_coarse-graining_condition}
    \Sigma_c(\bK, \wn) = \frac{N_c}{V_\text{BZ}} \int_\text{BZ} \! d\bk' \, \phi_\bK(\bk') \Sigma_{\text{DCA}^+}(\bk', \wn) .
\end{equation}
While the DCA lattice self-energy of Eq.~(\ref{eq:Sigma_DCA}) trivially satisfies this equation, the DCA$^+$ algorithm searches for a solution with continuous momentum dependence and inverts Eq.~(\ref{eq:self-energy_coarse-graining_condition}) in two steps.
First, the left-hand side, the cluster self-energy $\Sigma_c(\bK, \wn)$, is interpolated onto the full lattice Brillouin zone,
\begin{subequations}
\begin{align}
    \big\{\bK, \Sigma_c(\bK, \wn)\big\} &\rightarrow \big\{ \bk, \tilde{\Sigma}(\bk, \wn) \big\} , \\
    \text{with} \quad \tilde{\Sigma}(\bK, \wn) &= \Sigma_c(\bK, \wn) ,
\end{align}
\end{subequations}
and the result symmetrized to restore the point group symmetry of the lattice.
By replacing the cluster momentum $\bK$ with a general lattice momentum $\bk$, Eq.~(\ref{eq:self-energy_coarse-graining_condition}) becomes a convolution equation~\cite{Staar:2013ec},
\begin{equation}
\label{eq:self-energy_deconvolution}
    \tilde{\Sigma}(\bk, \wn) = \frac{N_c}{V_\text{BZ}} \int_\text{BZ} \! d\bk' \, \phi(\bk-\bk') \Sigma_{\text{DCA}^+}(\bk', \wn) ,
\end{equation}
where translational invariance,
\begin{equation}
    \phi_\bK(\bk) = \phi(\bk - \bK) ,
\end{equation}
and inversion symmetry,
\begin{equation}
    \phi(\bk - \bK) = \phi(\bK - \bk) ,
\end{equation}
allow the patch functions to be written as the convolution kernel.
The second step consists of inverting this convolution.
Staar~et~al.~\cite{Staar:2013ec} proposed the Richardson-Lucy deconvolution algorithm for solving this problem.

The DCA$^+$ lattice self-energy with full momentum resolution, $\Sigma_{\text{DCA}^+}(\bk, \wn)$, replaces the piecewise constant DCA approximation, $\Sigma_{\text{DCA}}(\bk, \wn)$, in the coarse-graining of the Green's function, Eq.~(\ref{eq:G_coarsegrained}).
If a QMC method is used that computes the new cluster self-energy $\Sigma_c(\bK, \wn)$ in a bare expansion~($\mathcal{G}_0$-formulation), DCA$^+$ requires the cluster-exclusion step, too,
\begin{equation}
\label{eq:cluster-exclusion_dca+}
    \mathcal{G}_0(\bK, \wn) = \left[ \bar{G}^{-1}(\bK, \wn) + \bar{\Sigma}(\bK, \wn) \right]^{-1} ,
\end{equation}
where $\bar{\Sigma}(\bK, \wn)$ is the coarse-grained average of the lattice self-energy $\Sigma_{\text{DCA}^+}(\bk, \wn)$,
\begin{equation}
\label{eq:Sigma_coarsegrained}
    \bar{\Sigma}(\bK, \wn) = \frac{N_c}{V_\text{BZ}} \int_\text{BZ} \! d\bk \, \phi_\bK(\bk) \Sigma_{\text{DCA}^+}(\bk, \wn) .
\end{equation}

\section{Failure of the DCA$^+$ algorithm}
\label{sec:failure_DCA+}

\begin{figure*}
    \includegraphics{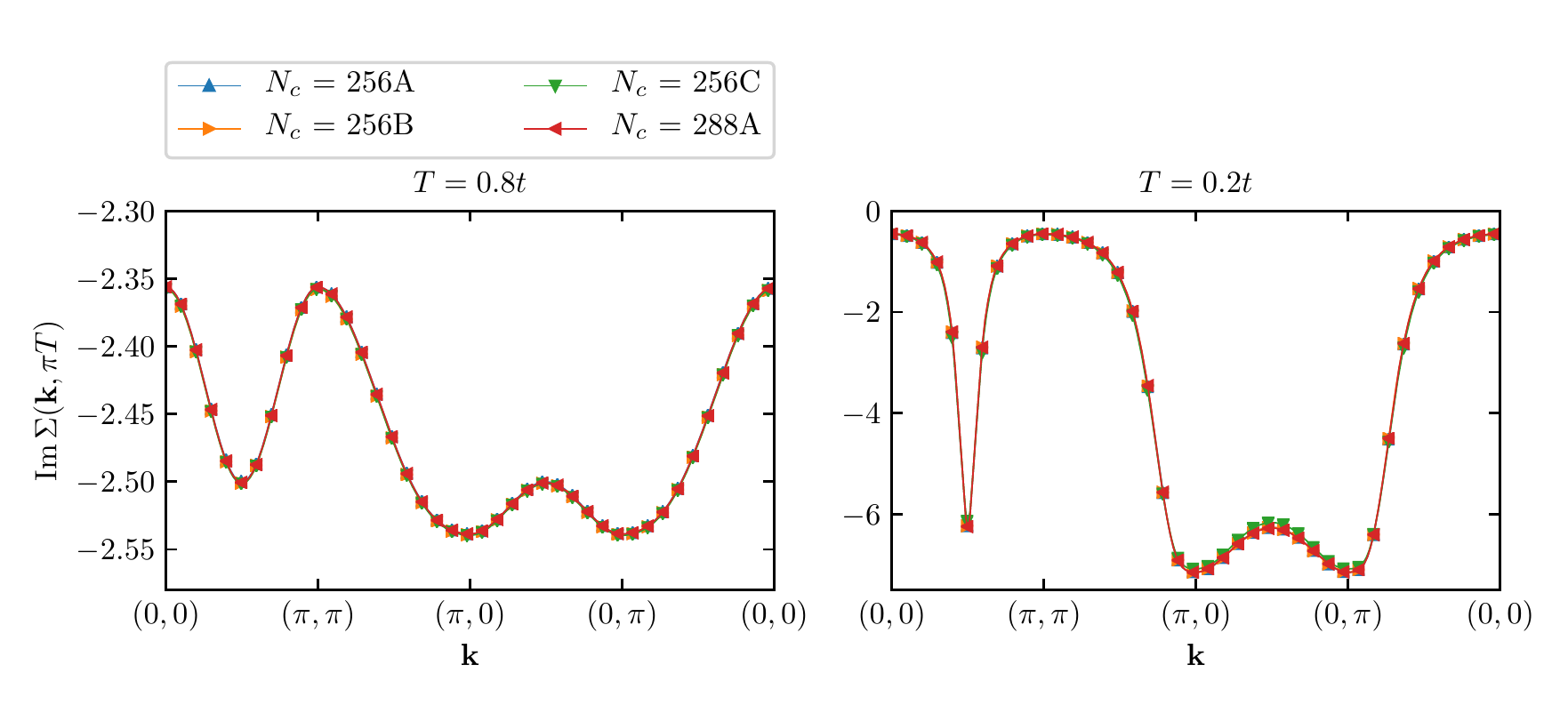}
    \caption{
        (Color online)
        Interpolated DCA results for the momentum dependence of the imaginary part of the self-energy at the lowest Matsubara frequency for the half-filled Hubbard model at $U = 7t$ and temperatures $T = 0.8t$~(left) and $T = 0.2t$~(right), respectively.
        The curves of the three different $N_c = 256$ clusters and the square $N_c = 288$ cluster match nearly perfectly, verifying convergence and cluster shape independence of the results.
    }
    \label{fig:DCA_Nc=256_288}
\end{figure*}

\begin{figure*}
    \includegraphics{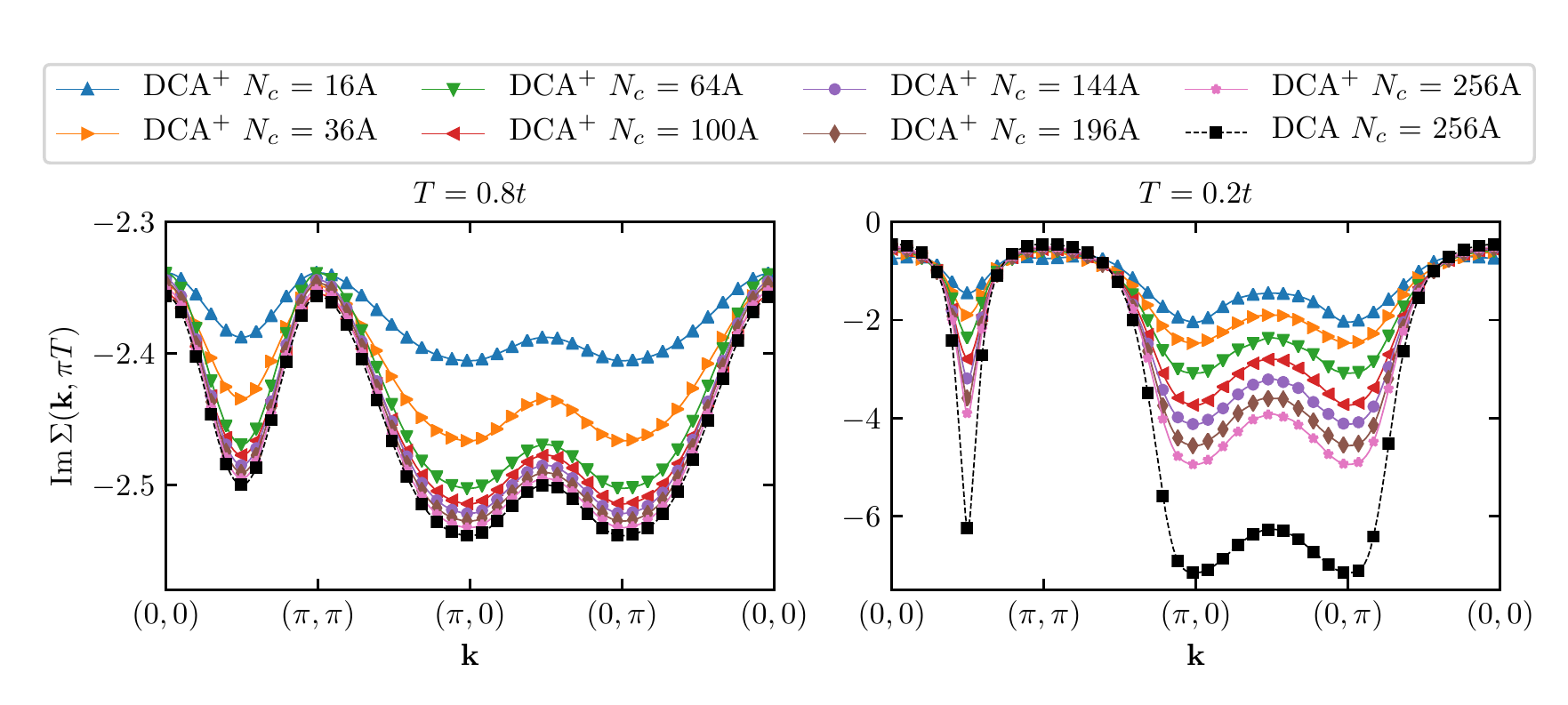}
    \caption{
        (Color online)
        Comparison of the momentum dependence of the imaginary part of the DCA$^+$ lattice self-energy at the lowest Matsubara frequency for various cluster sizes with the interpolated $N_c = 256A$ DCA result.
        The set of parameters is the same as in Fig.~\ref{fig:DCA_Nc=256_288}: $U = 7t$, $\langle n \rangle = 1$, and $T = 0.8t$~(left) and $T = 0.2t$~(right), respectively.
    }
    \label{fig:DCA+_vs_DCA_Nc=256_interpolated}
\end{figure*}

\begin{figure*}
    \includegraphics{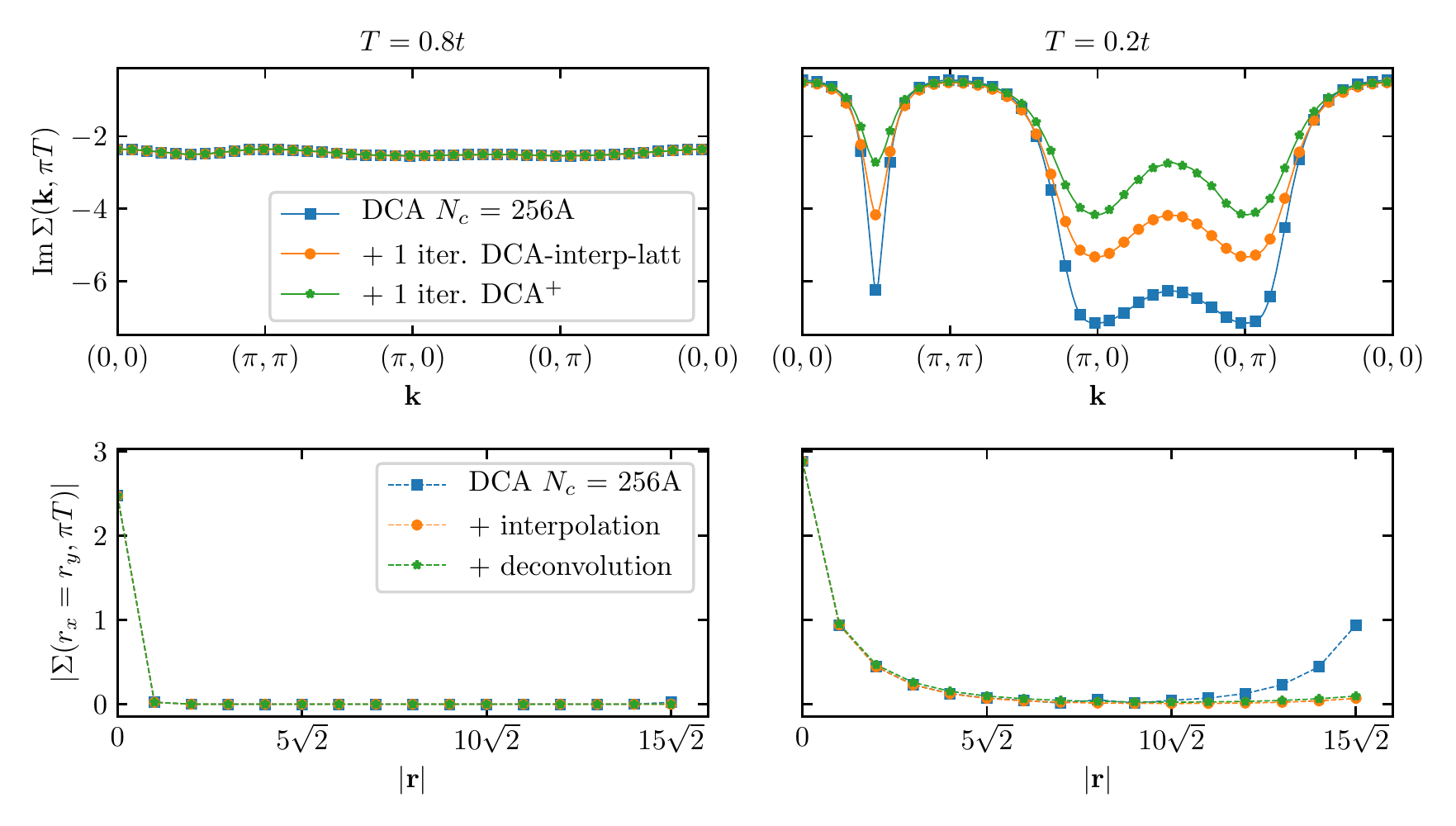}
    \caption{
        (Color online)
        Top: Momentum dependence of the imaginary part of the self-energy at the lowest Matsubara frequency for $U = 7t$, $\langle n \rangle = 1$, $N_c = 256A$, and $T = 0.8t$~(left) and $T = 0.2t$~(right), respectively.
        Performing only one DCA$^+$ iteration (green stars) or one {DCA-interp-latt} iteration (orange circles) on the converged DCA results (blue squares) yields a far more local self-energy at $T = 0.2t$.
        Bottom: Decay of the real space self-energy at the lowest Matsubara frequency for the same set of parameters.
        The DCA cluster self-energy (blue squares) is, by construction, periodic on the cluster.
        The interpolated (orange circles) and the deconvoluted (green stars) DCA cluster self-energy have the full momentum resolution of the lattice and are not periodic on the cluster.
    }
    \label{fig:DCA_converged_plus_X_self_energy}
\end{figure*}

\begin{figure*}
    \includegraphics{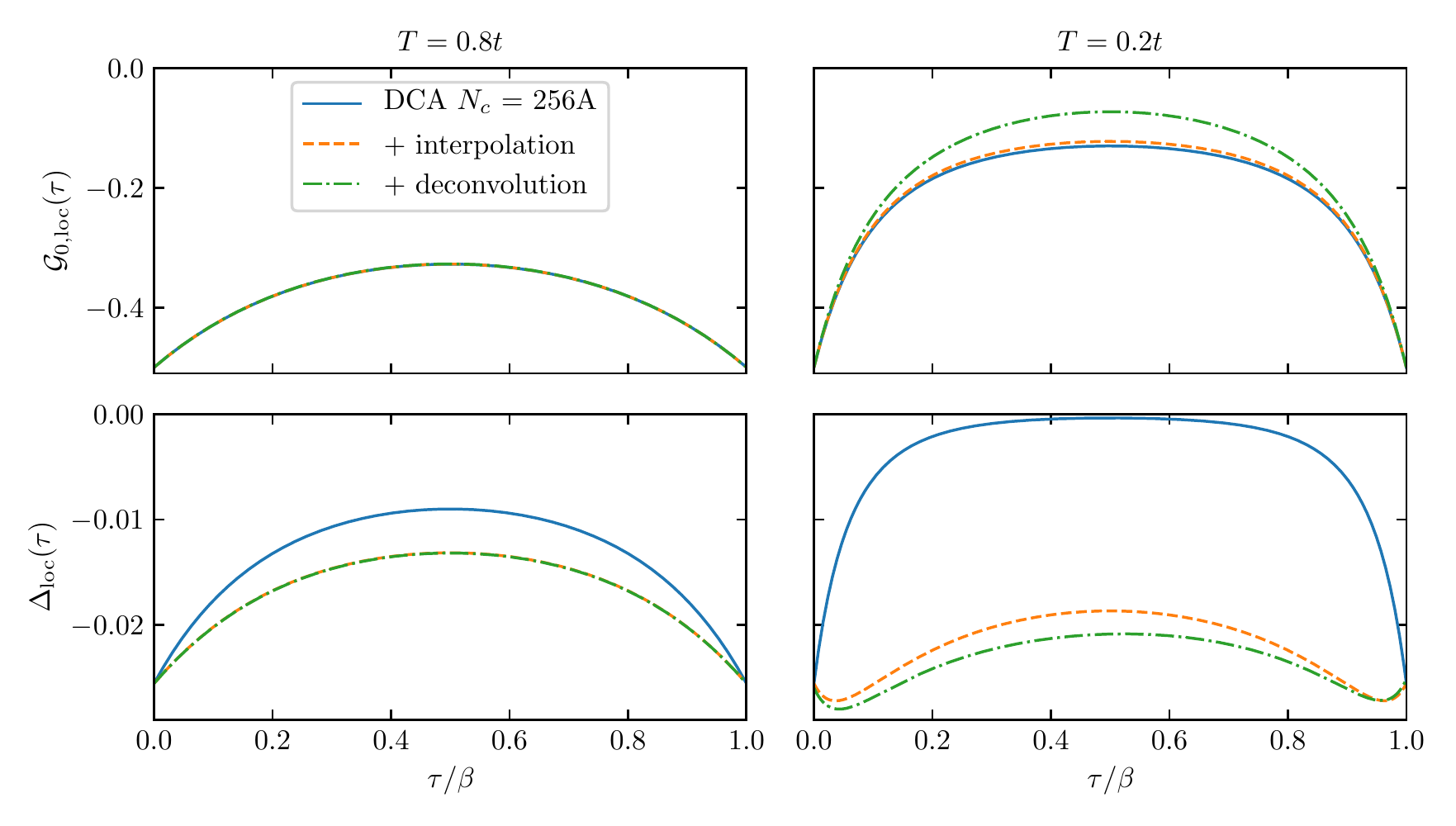}
    \caption{
        (Color online)
        Local bare cluster Green's function (top) and local hybridization function (bottom) in imaginary time for $U = 7t$, $\langle n \rangle = 1$, $N_c = 256A$, and $T = 0.8t$~(left) and $T = 0.2t$~(right), respectively.
        The solid blue lines correspond to the DCA results.
        The dashed orange lines represent results where the Green's function is coarse-grained with the interpolated DCA cluster self-energy.
        In the dash-dotted green results the Green's function is coarse-grained with the deconvoluted DCA cluster self-energy, and the cluster-exclusion step is done with the coarse-grained average of the latter.
    }
    \label{fig:DCA_converged_plus_X_cx_gf_and_hyb}
\end{figure*}

As a benchmark for the following analysis of the DCA$^+$ algorithm, Fig.\ref{fig:DCA_Nc=256_288} provides converged, cluster shape independent DCA results for the self-energy in the strongly coupled regime ($U = 7t, \langle n \rangle = 1)$ at two different temperatures.
Fig.~\ref{fig:DCA+_vs_DCA_Nc=256_interpolated} shows DCA$^+$ results for the same set of parameters and various cluster sizes.
At the higher temperature, $T = 0.8t$, the DCA benchmark self-energy is only weakly momentum dependent, and the DCA$^+$ results converge to the DCA benchmark as the cluster size is increased.
However, at the lower temperature, $T = 0.2t$, the DCA benchmark self-energy is strongly momentum dependent, and the DCA$^+$ results fail to converge to the DCA benchmark.
Even for a 256-site cluster, the DCA$^+$ self-energy is considerably smaller in magnitude and less momentum dependent than the DCA benchmark, in agreement with the observations by Vu{\v c}i{\v c}evi{\'c}~et~al.~\cite{Vucicevic:2018ev}.

To understand the problem in the DCA$^+$ algorithm that leads to the failure when the self-energy is strongly momentum dependent, we look at the effect of a single DCA$^+$ iteration on the converged DCA $N_c = 256A$ result.
The top right panel of Fig.~\ref{fig:DCA_converged_plus_X_self_energy} compares the resulting DCA$^+$ lattice self-energy with the interpolated DCA input at $T = 0.2t$ and proves two facts:
First, even when initialized with the quasi-exact self-energy, DCA$^+$ produces a quantitatively much smaller result at $T = 0.2t$.
Second, the failure happens within one iteration.
The same plot also shows results of a similar numerical experiment.
Instead of DCA$^+$, we perform one iteration of \emph{DCA with interpolated lattice self-energy ({DCA-interp-latt})}, where the Green's function is coarse-grained with the interpolated cluster self-energy.
Here, we observe a more local self-energy, too, although to a lesser extent than with DCA$^+$.

Besides the interacting part of the Hamiltonian, which is fixed, the bare cluster Green's function $\mathcal{G}_0$ is the only input to the QMC solver, which generates the new cluster self-energy.
Since DCA$^+$ and {DCA-interp-latt} already fail after one iteration, the problem must be caused by a difference in $\mathcal{G}_0$.
The top panels of Fig.~\ref{fig:DCA_converged_plus_X_cx_gf_and_hyb} compare the local part of the bare cluster Green's function in imaginary time between DCA, DCA$^+$, and {DCA-interp-latt}.
However, qualitative changes cannot be seen even at $T = 0.2t$.

A better picture is provided by the hybridization function $\Delta$, which is closely related to the bare cluster Green’s function $\mathcal{G}_0$ via
\begin{equation}
    \mathcal{G}_0(\bK, \wn) = \frac{1}{\wn + \mu - \bar{\epsilon}_\bK - \Delta(\bK, \wn)} ,
\end{equation}
where $\bar{\epsilon}_\bK$ is the coarse-grained average of the dispersion $\epsilon_\bk$.
The bottom panels of Fig.~\ref{fig:DCA_converged_plus_X_cx_gf_and_hyb} show the local part of the hybridization function in imaginary time.
Already at $T = 0.8t$, we see that both DCA$^+$ and {DCA-interp-latt} yield a larger, in absolute value, hybridization function than standard DCA.
The behavior in the case of DCA$^+$ can be understood from the following explicit form of the DCA$^+$ hybridization function (derivation in Appendix~\ref{app:DCA+_hyb_function}),
\begin{equation}
    \Delta(\bK, \wn) = \frac{\langle \left[\delta \epsilon(\bk) + \delta \Sigma(\bk, \wn)\right]^2 G(\bk, \wn) \rangle_\bK}{1 + \langle \left[\delta \epsilon(\bk) + \delta \Sigma(\bk, \wn)\right] G(\bk ,\wn) \rangle_\bK} .
\end{equation}
We can see that the difference between the lattice and the cluster self-energy, $\delta \Sigma(\mathbf{k}) = \Sigma_{\text{DCA}^+}(\bk, \wn) - \bar{\Sigma}(\bK, \wn)$, appears like an additional hopping amplitude and consequently increases the hybridization of the cluster with the bath.
This explains why a more metallic, less correlated solution with a smaller self-energy is found when the cluster self-energy differs from the continuous lattice self-energy.

The discrepancy to standard DCA becomes even more dramatic at $T = 0.2t$.
In addition to an even larger difference in value, the local hybridization functions of DCA$^+$ and {DCA-interp-latt} possess local minima, which correspond to causality violations of their second derivatives, as the latter become positive.
These causality violations in the case of DCA$^+$ were observed by Vu{\v c}i{\v c}evi{\'c}~et~al.~\cite{Vucicevic:2018ev}, as well.
While the DCA self-consistency loop in the $\mathcal{G}_0$-formulation was proven to be causal~\cite{Hettler:2000es}, the proof does neither apply to DCA$^+$ nor to {DCA-interp-latt}, as these results confirm.

The unphysicality of both the DCA$^+$ and the {DCA-interp-latt} hybridization function must originate from a problem in the definition of the bare cluster Green's function, i.e., the cluster-exclusion step.
As explained in Sec.~\ref{sec:methods}, its purpose is to remove the self-energy contributions on the cluster (i.e., the cluster self-energy) from the coarse-grained Green's function.
However, only in the standard DCA algorithm does the lattice self-energy that is used in the coarse-graining, in a piecewise constant continuation, match the cluster self-energy.
DCA$^+$ and {DCA-interp-latt} employ a deconvolution and an interpolation of the cluster self-energy, respectively.
A comparison of the three different representations of the lattice self-energy is most illustrative in real space, as depicted by the bottom panels of Fig.~\ref{fig:DCA_converged_plus_X_self_energy}.
While the DCA cluster self-energy is, by construction, periodic on the real space cluster, the continuous lattice self-energies are not.
Furthermore, comparing the bottom panels with the top panels reveals that failure of the methods with continuous lattice self-energy happens when the self-energy is long-ranged so that the periodicity gives the DCA cluster self-energy an increasing tail.
Hence, the fundamental problem of DCA$^+$ and {DCA-interp-latt} can be traced to the mismatch between the non-periodic lattice self-energy added in the coarse-graining of the Green's function and the periodic cluster self-energy removed in the cluster-exclusion step.

\begin{figure}
    \includegraphics{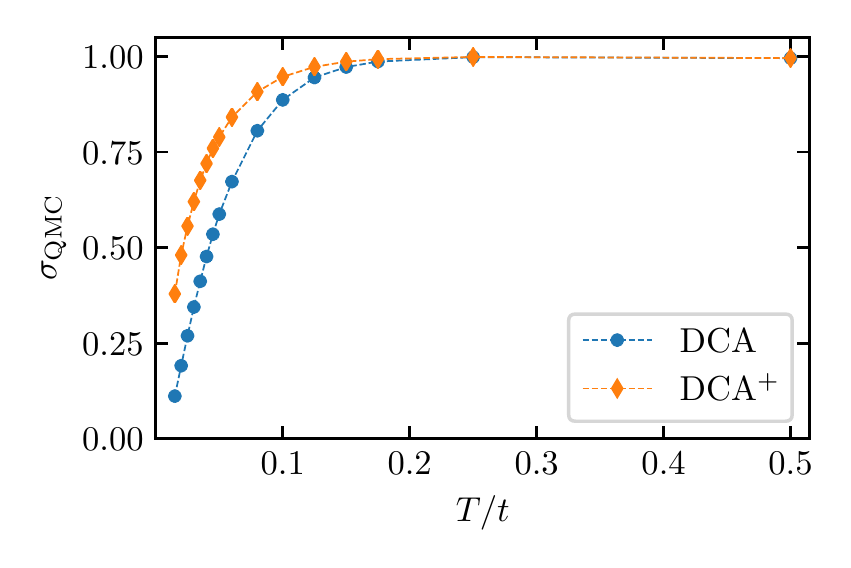}
    \caption{
        (Color online)
        Average QMC sign $\sigma_\text{QMC}$ vs. temperature $T$ for the 10\% hole-doped Hubbard model with $U = 4t$ and cluster size $N_c = 24$.
    }
    \label{fig:DCA_vs_DCA+_U=4_sign}
\end{figure}

\begin{figure*}
    \includegraphics{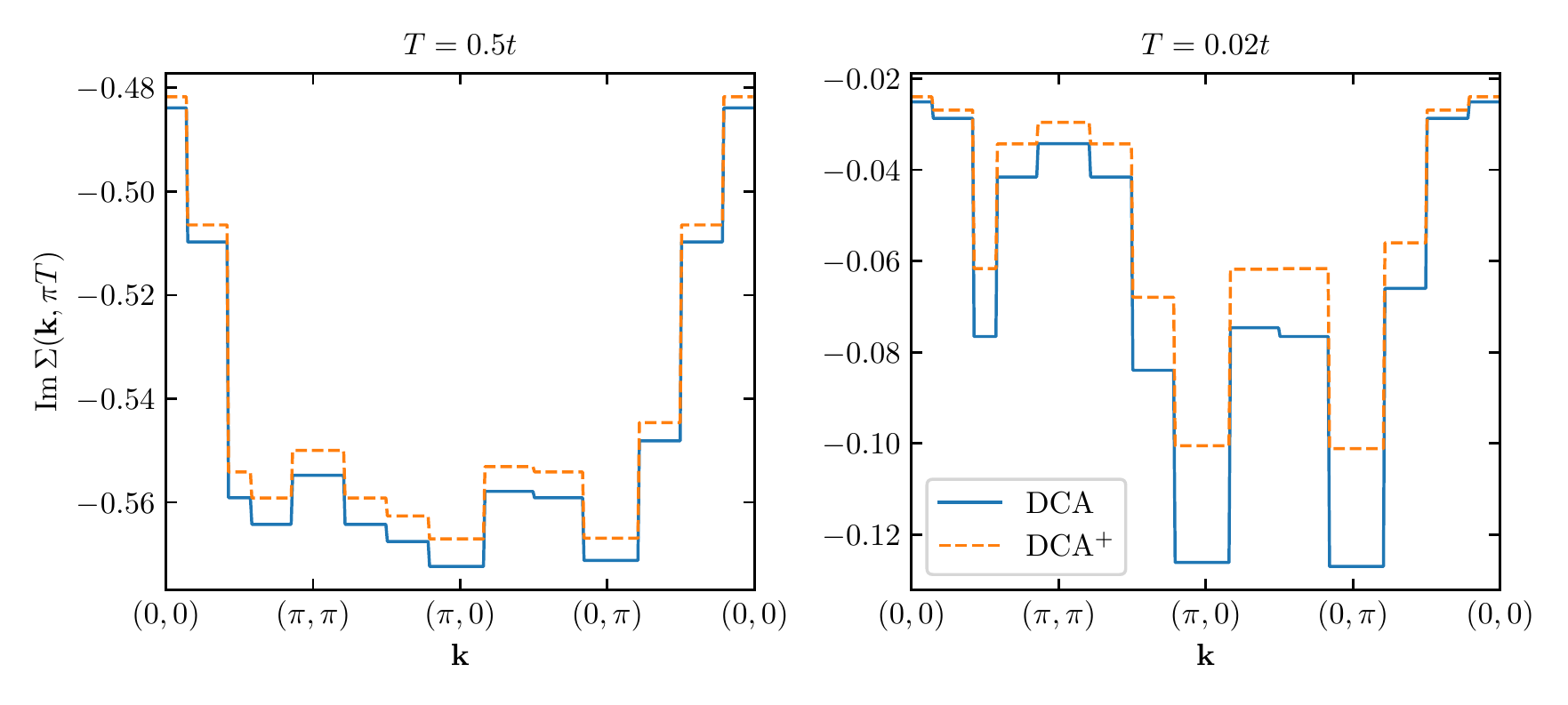}
    \caption{
        (Color online)
        DCA and DCA$^+$ results for the momentum dependence of the imaginary part of the self-energy at the lowest Matsubara frequency for the 10\% hole-doped Hubbard model with $U = 4t$, $N_c = 24$, and temperatures $T = 0.5t$~(left) and $T = 0.02t$~(right), respectively.
    }
    \label{fig:DCA_vs_DCA+_U=4_self_energy}
\end{figure*}

To conclude this section, we want to comment on the reduction of the fermionic sign problem in DCA$^+$, as demonstrated by Fig.~\ref{fig:DCA_vs_DCA+_U=4_sign} for the 10\% hole-doped Hubbard model at $U = 4t$ and cluster size $N_c = 24$.
In Ref.~\cite{Staar:2013ec}, the larger average QMC sign in DCA$^+$ compared to standard DCA was argued to arise from the removal of artificial long-range correlations introduced by the jump discontinuities of the DCA self-energy.
However, based on the insights from this section, we draw a different conclusion.
Fig.~\ref{fig:DCA_vs_DCA+_U=4_self_energy} indicates that the reduction of the sign problem in the DCA$^+$ algorithm merely comes from the fact that it produces a less correlated result with smaller self-energy at low temperatures.

\section{DCA with post-interpolation: Application to the two-dimensional Hubbard model}
\label{sec:DCA_with_post_interpolation_application}

The fundamental problem of the DCA$^+$ algorithm and the resulting failure when the self-energy is strongly momentum dependent make it a poor choice in the strongly correlated regime at low doping.
Yet, resorting to the standard DCA algorithm means suffering again from its cluster shape dependence, poor convergence with cluster size, and broken lattice symmetries if clusters are used that do not possess the full point group symmetry of the lattice.
Interpolating the cluster self-energy clearly has the potential to help with these issues.
However, as discussed in Sec.~\ref{sec:failure_DCA+}, interpolating the cluster self-energy within the self-consistency loop ({DCA-interp-latt}) is plagued by the same problem as the DCA$^+$ algorithm.
For this reason, we propose a different, practical approach, \emph{DCA with post-interpolation}, in which we interpolate and symmetrize only the final DCA cluster self-energy.
The details of the implementation are described in Appendix~\ref{app:interpolation_and_symmetrization_of_the_self_energy}.

\begin{figure*}
    \includegraphics{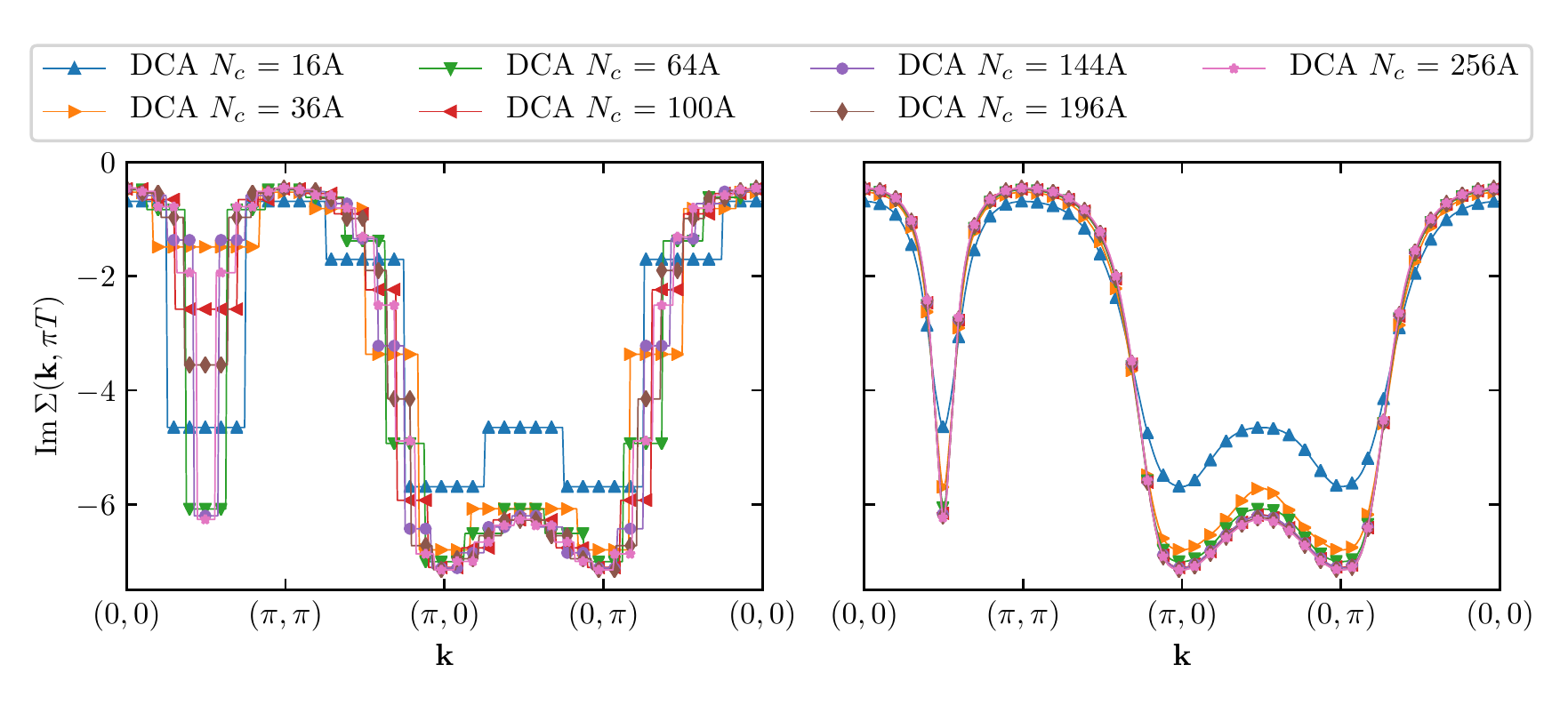}
    \caption{
        (Color online)
        Momentum dependence of the piecewise constant (left) and the interpolated and symmetrized (right) DCA self-energy at the lowest Matsubara frequency for the half-filled Hubbard model at $U = 7t$ and $T = 0.2t$ for various $[[2L, 0], [0, 2L]]$ clusters.
    }
    \label{fig:DCA_cluster_size_convergence_square_clusters}
\end{figure*}

DCA with post-interpolation yields a lattice self-energy with full momentum resolution.
To show the effect of this, Fig.~\ref{fig:DCA_cluster_size_convergence_square_clusters} compares the piecewise constant DCA lattice self-energy with the interpolated and symmetrized result for various $[[2L, 0], [0, 2L]]$ clusters.
As $\bk = (\pi/2, \pi/2)$ is only a cluster momentum for even $L$, the piecewise constant DCA lattice self-energy shows poor convergence here.
The interpolation resolves this and reveals the systematic convergence with cluster size.
In particular, one sees that for this parameter set, the interpolated DCA results for all momenta are already well converged for clusters larger than 36 sites.

\begin{figure*}
    \includegraphics{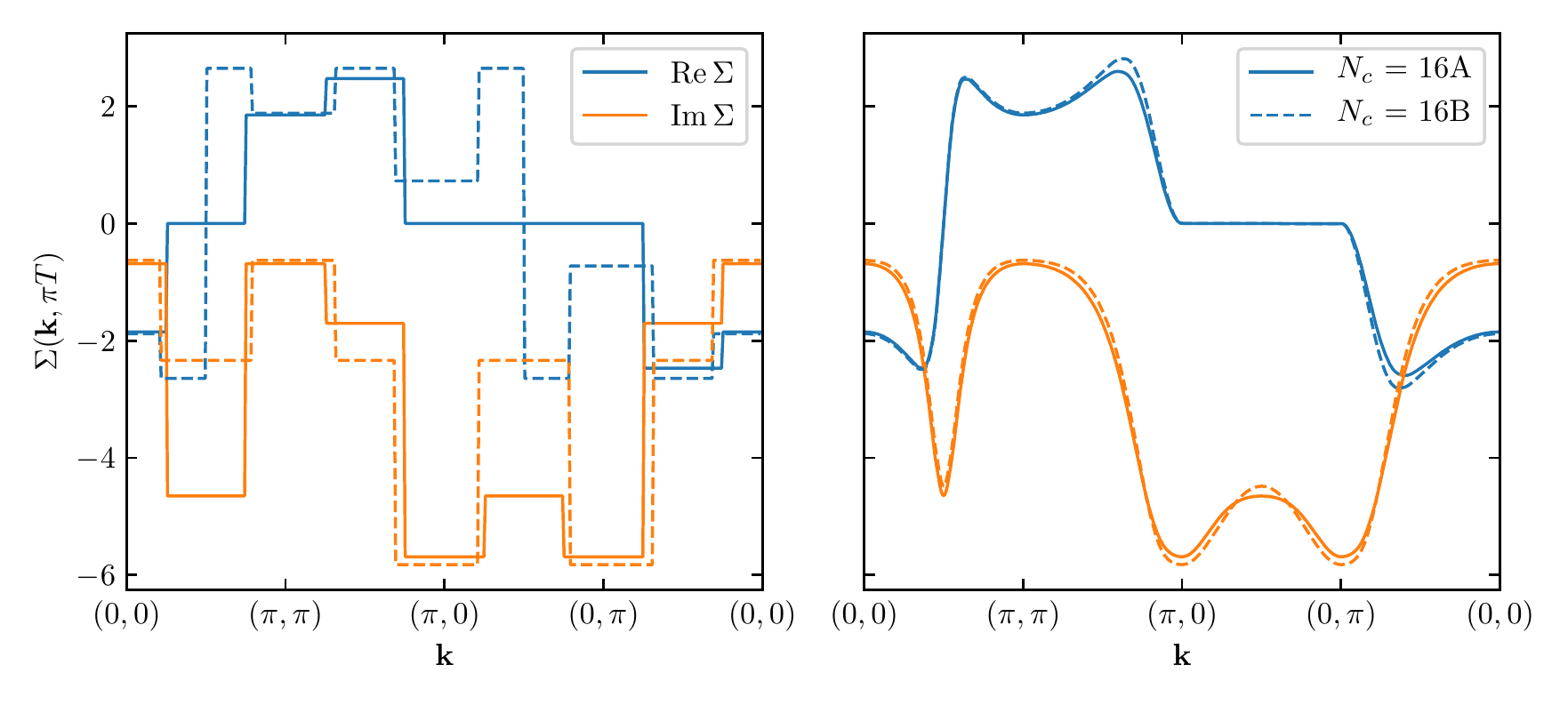}
    \caption{
        (Color online)
        Momentum dependence of the piecewise constant (left) and the interpolated and symmetrized (right) DCA self-energy at the lowest Matsubara frequency for the half-filled Hubbard model at $U = 7t$ and $T = 0.2t$ for two different 16-site clusters.
    }
    \label{fig:DCA_cluster_shape_dependence}
\end{figure*}

The post-interpolation approach can also reveal cluster shape independence of the DCA results as demonstrated in Fig.~\ref{fig:DCA_cluster_shape_dependence}.
The piecewise constant DCA lattice self-energy shows large deviations between the 16A and 16B clusters around $\bk = (\pi/2, \pi/2)$ due to the different shape and position of the coarse-graining patches.
The interpolated and symmetrized results of the two clusters are in remarkably good agreement.

\begin{figure*}
    \includegraphics{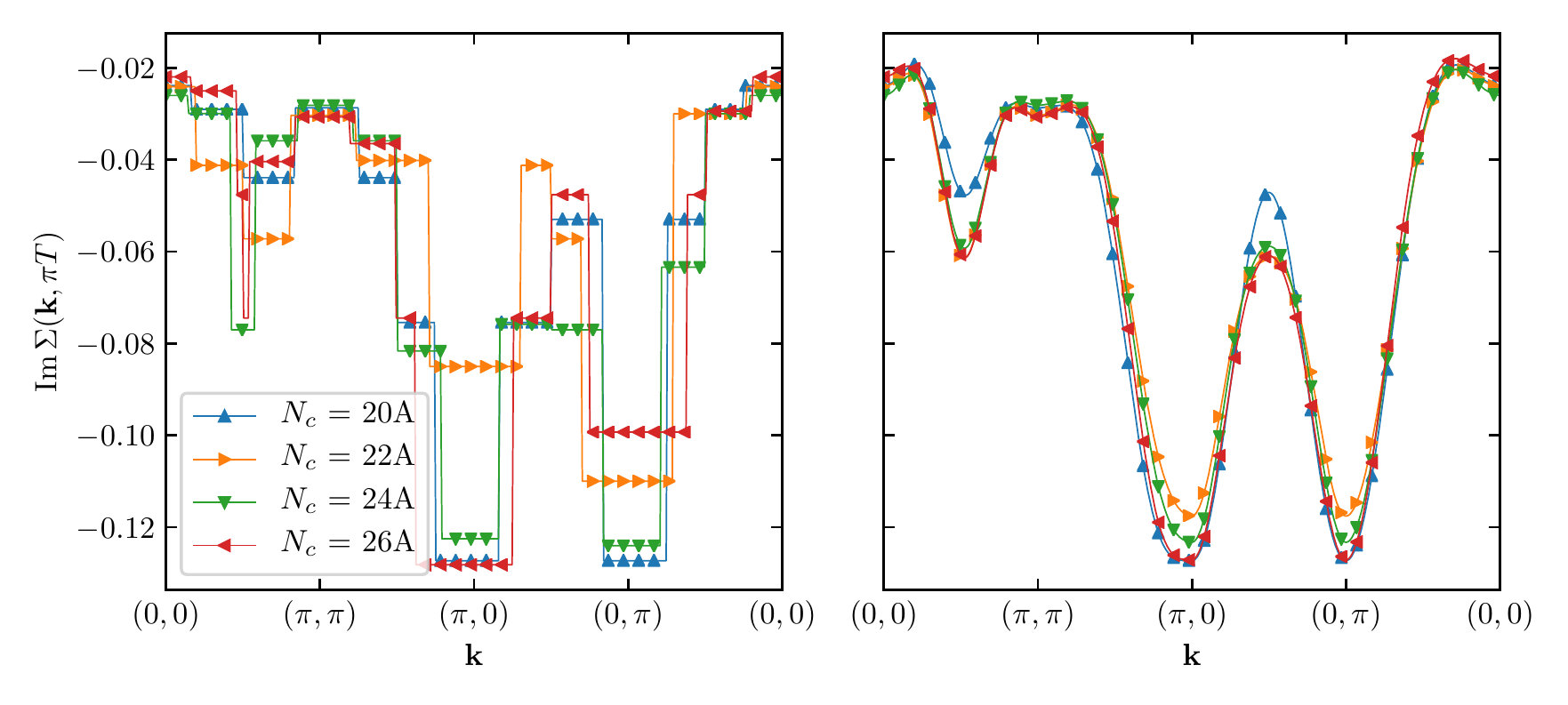}
    \caption{
        (Color online)
        Momentum dependence of the piecewise constant (left) and the interpolated and symmetrized (right) DCA self-energy at the lowest Matsubara frequency for the 10\% hole-doped Hubbard model at $U = 4t$ and $T = 0.02t$ for several similar sized clusters.
    }
    \label{fig:DCA_cluster_size_convergence_converged}
\end{figure*}

Fig.~\ref{fig:DCA_cluster_size_convergence_converged} shows data for the 10\% hole-doped Hubbard model at $U = 4t$ and $T = 0.02t$ for cluster sizes between 20 and 26.
For this set of parameters, the fermionic sign problem of the QMC solver renders calculations on larger clusters such as $N_c = 36$ or $N_c = 64$ impossible.
None of the clusters used in Fig.~\ref{fig:DCA_cluster_size_convergence_converged} obey the full point group symmetry of the square lattice, which causes the piecewise constant DCA lattice self-energy to vary significantly between them.
After interpolation and symmetrization with the full point group symmetry of the lattice, all clusters show similar results, indicating convergence with cluster size.

As the DCA$^+$ algorithm~\cite{Staar:2014gz}, DCA with post-interpolation can be extended to the two-particle level for the study of phase transitions.
To determine the superconducting transition temperature $T_c$, for example, we need to compute the eigenvalues $\lambda_\alpha$ of the Bethe-Salpeter kernel~\cite{Maier:2006gr},
\begin{equation}
    -\frac{T}{N} \sum_{k'} \Gamma^{pp}(k, k') G(k') G(-k') g_\alpha(k') = \lambda_\alpha g_\alpha(k) ,
\end{equation}
where $k \equiv (\bk, \wn)$, $\Gamma^{pp}(k, k')$ is the irreducible particle-particle vertex on the lattice, and $G(\bk, \wn)$ is the lattice Green's function computed from the interpolated and symmetrized cluster self-energy.
Analog to the procedure for the self-energy, we compute the lattice vertex $\Gamma^{pp}(k, k')$ from an interpolation of the cluster vertex $\Gamma_c^{pp}(K, K')$ and symmetrize the result according to the point group symmetry of the lattice.

\begin{figure}
    \includegraphics{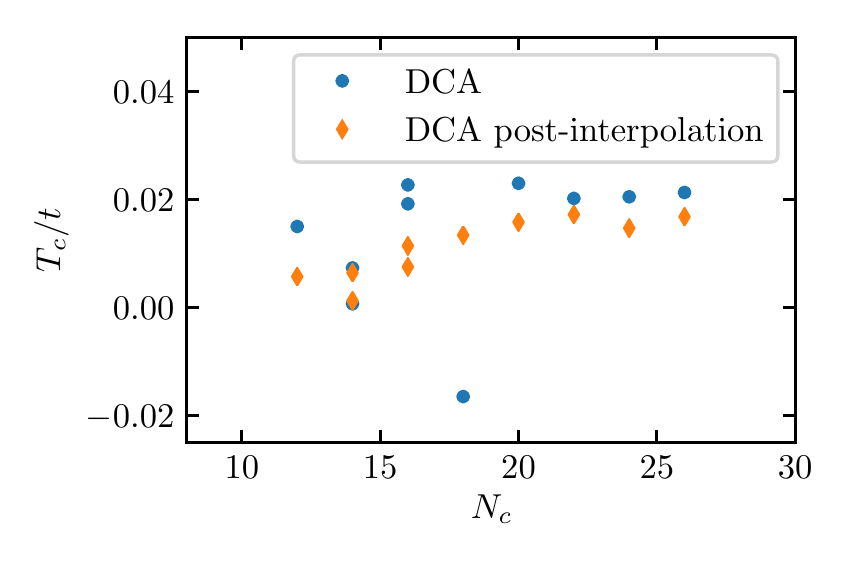}
    \caption{
        (Color online)
        Superconducting transition temperature $T_c$ vs. cluster size $N_c$ for the 10\% hole-doped Hubbard model with $U = 4t$.
        The two data points at $N_c = 14$ and $N_c = 16$ correspond to the 14A and 14B clusters and the 16A and 16B clusters, respectively.
    }
    \label{fig:Tc_vs_Nc_U=4_p=0}
\end{figure}

Fig.~\ref{fig:Tc_vs_Nc_U=4_p=0} shows the cluster size dependence of $T_c$ for the 10\% hole-doped Hubbard model with $U = 4t$, obtained with the standard DCA and DCA with post-interpolation~\footnote{The irreducible particle-particle vertex $\Gamma^{pp}(k, k', q)$ was computed for fixed $\mathbf{q} = 0$ and bosonic Matsubara frequency $\nu_m = 0$ and for 16 positive fermionic Matsubara frequencies.}.
Most notable is the $N_c = 18$ outlier in the DCA results, which is caused by an unfavorable choice of cluster momenta in the first Brillouin zone:
The set of cluster momenta lacks $\bk = (\pi, 0)$ and $\bk' = (0, \pi)$, between which the pairing interaction $\Gamma^{pp}(k, k')$ is the largest~\cite{Maier:2006gr}, and thus $T_c$ is vastly underestimated.
This is resolved in DCA with post-interpolation, where $T_c$ is computed from an irreducible particle-particle vertex $\Gamma^{pp}(k, k')$ that has been interpolated onto the full lattice Brillouin zone, including $\bk = (\pi, 0)$ and $\bk' = (0, \pi)$.
Besides the elimination of the $N_c = 18$ outlier, the results also show that the post-interpolation approach leads to a more controlled convergence with cluster size.

\section{Conclusions and outlook}
\label{sec:conclusions}

We have uncovered the fundamental problem of the DCA$^+$ algorithm in the $\mathcal{G}_0$-formulation.
The computation of the bare cluster Green's function $\mathcal{G}_0$ is affected by a mismatch between the continuous lattice self-energy entering the coarse-grained Green's function and the periodic cluster self-energy that is subtracted.
We have shown that, as a result, DCA$^+$ fails in the regime where the self-energy is long-ranged and the periodic cluster self-energy differs substantially from the continuous lattice self-energy.
This difference is absorbed in and increases the magnitude of the hybridization function, which describes the mean-field coupling of the cluster to the rest of the system, thus resulting in a more metallic and less correlated solution.
Moreover, we have argued that the reduction of the sign problem in DCA$^+$ is a consequence of this failure.

It is interesting to note the parallels between DCA$^+$ and the periodized cluster dynamical mean-field theory~(PCDMFT)~\cite{Biroli:2004jg} with respect to their formulation and failure.
The PCDMFT is a generalization of the CDMFT that, like DCA$^+$, replaces the cluster self-energy in the computation of the Green's function with the lattice self-energy.
Again, this leads to a mismatch in the computation of the bare cluster Green's function.
Vu{\v c}i{\v c}evi{\'c}~et~al.~\cite{Vucicevic:2018ev} showed that the PCDMFT indeed fails in a similar region of the phase diagram as DCA$^+$ and that the failure coincides with causality violations in the hybridization function, too.
However, Biroli~et~al.~\cite{Biroli:2004jg} provided a proof that the PCDMFT is causal in the $\Phi$-formulation.
This means that the problem in the PCDMFT must be an artifact of the $\mathcal{G}_0$-formulation.
We expect the same to be true for the DCA$^+$ algorithm.
Hence, a challenge is the development of alternative iterative procedures to solve the self-consistency equations in the $\Phi$-formulation.
One has to be aware, however, that the recently discovered multivaluedness of the Luttinger-Ward functional~\cite{Kozik:2015jq, Gunnarsson:2017co} could be an issue in such iterative procedures.

Another consequence of our findings is that we need to extend the list of requirements for a satisfactory quantum cluster theory for correlated systems.
We should add that quantum cluster algorithms in the $\mathcal{G}_0$-formulation should use the same self-energy both in the coarse-graining of the Green's function and for the computation of the bare cluster Green's function.

As the standard DCA is the only algorithm known to date that satisfies all requirements, we have proposed a practical approach for introducing continuity in the DCA.
By interpolating and symmetrizing the final DCA self-energy and irreducible vertex, we obtain results with full momentum resolution that preserve the point group symmetry of the lattice.
While the approach has no effect on the severity of the fermionic sign problem, it reduces the cluster shape dependence of the DCA and improves the convergence with cluster size.

\begin{acknowledgments}
    All simulations were done using the DCA++ code~\cite{Hahner:2020cs}.
    The work of T.A.M. was supported by the Scientific Discovery through Advanced Computing~(SciDAC) program funded by U.S. Department of Energy, Office of Science, Advanced Scientific Computing Research and Basic Energy Sciences, Division of Materials Sciences and Engineering.
    This research used resources of the Oak Ridge Leadership Computing Facility~(OLCF) awarded by the INCITE program and of the Swiss National Supercomputing Center~(CSCS).
    OLCF is a DOE Office of Science User Facility supported under Contract DE-AC05-00OR22725.
\end{acknowledgments}

\appendix

\section{Derivation of the DCA$^+$ hybridization function}
\label{app:DCA+_hyb_function}

To simplify the notation, we denote the coarse-grained average of a function $f(\bk)$ by
\begin{equation}
    \langle f \rangle_\bK = \frac{N_c}{V_\text{BZ}} \int_\text{BZ} \! d\bk \, \phi_\bK(\bk) f(\bk) ,
\end{equation}
represent the dependence on the cluster momentum by a subscript $\bK$, and omit the frequency argument $\wn$ of Green's functions, self-energies, and the hybridization function.

First, we split the lattice dispersion $\epsilon(\bk)$ and the DCA$^+$ lattice self-energy $\Sigma_{\text{DCA}^+}(\bk)$ in their coarse-grained part and a correction term,
\begin{align}
    \epsilon(\bk) &=  \langle \epsilon \rangle_\bK + \delta \epsilon(\bk) , \\
    \Sigma_{\text{DCA}^+}(\bk) &= \langle \Sigma \rangle_\bK + \delta \Sigma(\bk) ,
\end{align}
where by construction the DCA$^+$ algorithm satisfies
\begin{equation}
    \langle \Sigma \rangle_\bK \equiv \bar{\Sigma}_\bK = \Sigma_{c,\bK} .
\end{equation}
As in the DCA, we define the hybridization function $\Delta_\bK$ via the bare cluster Green’s function,
\begin{equation}
    \mathcal{G}_{0, \bK} = \frac{1}{z + \mu - \langle \epsilon \rangle_\bK - \Delta_\bK} .
\end{equation}
Next, we introduce $g_\bK = \left[ {z + \mu - \langle \epsilon \rangle_\bK - \Sigma_{c, \bK}} \right]^{-1}$, which allows us to write the coarse-grained Green's function $\bar{G}_\bK \equiv \langle G \rangle_\bK$ in terms of the hybridization function $\Delta_\bK$,
\begin{subequations}
\begin{align}
    \langle G \rangle_\bK
        &= \frac{1}{\mathcal{G}_{0, \bK}^{-1} - \Sigma_{c, \bK}} \\
        &= \frac{1}{g_\bK^{-1} - \Delta_\bK} . \label{eq:G_bar_1}
\end{align}
\end{subequations}
We can also express the lattice Green's function $G(\bk)$ in terms of $g_\bK$,
\begin{subequations}
\begin{align}
    G(\bk)
        &= \frac{1}{z + \mu - \epsilon(\bk) - \Sigma_{\text{DCA}^+}(\bk) } \\
        &= \frac{1}{g_\bK^{-1} - \delta \epsilon(\bk) - \delta \Sigma(\bk)} ,
\end{align}
\end{subequations}
then write it in a Dyson equation as
\begin{equation}
\label{eq:G_k_1}
    G(\bk) = g_\bK + g_\bK \left[\delta \epsilon(\bk) + \delta \Sigma(\bk) \right] G(\bk) ,
\end{equation}
and iterate once,
\begin{align}
    G(\bk) = \, &g_\bK + g_\bK \left[\delta \epsilon(\bk) + \delta \Sigma(\bk)\right] \nonumber \\
        &\times \left\{ g_\bK + g_\bK \left[\delta \epsilon(\bk) + \delta \Sigma(\bk)\right] G(\bk) \right\} . \label{eq:G_k_2}
\end{align}
Taking the coarse-grained average of Eqs.~(\ref{eq:G_k_1}) and (\ref{eq:G_k_2}), respectively, we obtain
\begin{subequations}
\begin{align}
    \langle G \rangle_\bK &= g_\bK + g_\bK \langle \left[\delta \epsilon(\bk) + \delta \Sigma(\bk)\right] G(\bk) \rangle_\bK , \label{eq:G_bar_2} \\
    \langle G \rangle_\bK &= g_\bK + g_\bK^2 \langle \left[\delta \epsilon(\bk) + \delta \Sigma(\bk)\right]^2 G(\bk) \rangle_\bK ,
\end{align}
\end{subequations}
where we used $\langle \delta \epsilon \rangle_\bK = \langle \delta \Sigma \rangle_\bK = 0$.
By equating the last two expression for $\langle G \rangle_\bK$ and solving for $g_\bK$, we find
\begin{equation}
\label{eq:g_K}
    g_\bK = \frac{\langle \left[\delta \epsilon(\bk) + \delta \Sigma(\bk)\right] G(\bk) \rangle_\bK}{\langle \left[\delta \epsilon(\bk) + \delta \Sigma(\bk)\right]^2 G(\bk) \rangle_\bK} .
\end{equation}
Finally, we use Eqs.~(\ref{eq:G_bar_1}) and (\ref{eq:G_bar_2}) to solve for $\Delta_\bK$,
\begin{equation}
    \Delta_\bK = g_\bK^{-1} \times \frac{\langle \left[\delta \epsilon(\bk) + \delta \Sigma(\bk)\right] G(\bk) \rangle_\bK}{1 + \langle \left[\delta \epsilon(\bk) + \delta \Sigma(\bk)\right] G(\bk) \rangle_\bK} ,
\end{equation}
where we can replace $g_\bK$ with Eq.~(\ref{eq:g_K}) to obtain the final result,
\begin{equation}
\label{eq:Delta_K}
    \Delta_\bK = \frac{\langle \left[\delta \epsilon(\bk) + \delta \Sigma(\bk)\right]^2 G(\bk) \rangle_\bK}{1 + \langle \left[\delta \epsilon(\bk) + \delta \Sigma(\bk)\right] G(\bk) \rangle_\bK} .
\end{equation}

\section{Interpolation and symmetrization of the self-energy}
\label{app:interpolation_and_symmetrization_of_the_self_energy}

The interpolation procedure we apply to the self-energy is the same as described in the original DCA$^+$ paper, Ref.~\cite{Staar:2013ec}.
Instead of directly interpolating the self-energy, we first smooth it out with the following transformation,
\begin{equation}
\label{eq:alpha_trafo}
    \sigma(\bK, \wn) = \left[ \Sigma(\bK, \wn) - \text{sgn}(\omega_n) i \alpha \right]^{-1} , \quad \alpha > 0 .
\end{equation}
The smooth function $\sigma(\bK, \wn)$ is then interpolated onto the reciprocal lattice using Wannier interpolation,
\begin{subequations}
\begin{align}
    \sigma(\bR_c, \wn) &= \frac{1}{N_c} \sum_{\bK} e^{-i \bR_c \cdot \bK} \sigma(\bK, \wn) , \\
    \tilde{\sigma}(\bk, \wn) &= \sum_{\bR_c} e^{i \bR_c \cdot \bk} \sigma(\bR_c, \wn) w(\bR_c) .
\end{align}
\end{subequations}
Here, $\bR_c$ are \emph{centered} real space cluster vectors with corresponding weights $w(\bR_c)$.
To obtain the interpolated self-energy $\tilde{\Sigma}(\bk, \wn)$, the inverse transformation to Eq.~(\ref{eq:alpha_trafo}) has to be applied,
\begin{equation}
    \tilde{\Sigma}(\bk, \wn) = \tilde{\sigma}^{-1}(\bk, \wn) + \text{sgn}(\omega_n) i \alpha .
\end{equation}
Finally, $\tilde{\Sigma}(\bk, \wn)$ is symmetrized to restore the full point group symmetry of the lattice,
\begin{equation}
    \tilde{\Sigma}_\text{symm}(\bk, \wn) = \frac{1}{N_{\hat{S}}} \sum_{\hat{S}} \tilde{\Sigma}(\hat{S}(\bk), \wn) .
\end{equation}

\vspace{12mm}

\bibliography{refs.bib}

\end{document}